\documentclass[twocolumn]{27onr_art}
\usepackage{ulem}
\usepackage{harvard}
\usepackage{graphicx}
\usepackage{epsfig}
\usepackage{multirow}

\pdfoutput = 1

\def\be{\begin{eqnarray}}
\def\ee{\end{eqnarray}}
\def\benl{\begin{eqnarray*}}
\def\eenl{\end{eqnarray*}}

\newcommand{\nwc}{\newcommand}
\nwc{\bm}{\boldmath}
\nwc{\m}{\mbox}
\nwc{\ubm}{\unboldmath}
\nwc{\bmU}{\m{\bm$U$\ubm}}
\nwc{\bmX}{\m{\bm$X$\ubm}}
\nwc{\bmu}{\m{\bm$u$\ubm}}
\nwc{\bmx}{\m{\bm$x$\ubm}}
\nwc{\bmz}{\m{\bm$z$\ubm}}
\nwc{\bmv}{\m{\bm$v$\ubm}}
\nwc{\bmw}{\m{\bm$w$\ubm}}
\nwc{\bmW}{\m{\bm$W$\ubm}}
\nwc{\bmn}{\m{\bm$n$\ubm}}
\nwc{\bmG}{\m{\bm$G$\ubm}}
\nwc{\bmF}{\m{\bm$F$\ubm}}
\nwc{\bmI}{\m{\bm$I$\ubm}}
\nwc{\bmN}{\m{\bm$N$\ubm}}
\nwc{\bmP}{\m{\bm$P$\ubm}}
\nwc{\bmcalP}{\m{\bm $\cal P$\ubm}}
\nwc{\bmV}{\m{\bm$V$\ubm}}
\nwc{\bmS}{\m{\bm$S$\ubm}}

\begin{document}

%
%

\title{A comparison of full-scale experimental measurements and computational predictions of the transom-stern wave of the R/V Athena I}

\author{Donald C. Wyatt$^1$, Thomas C. Fu$^2$, Genevieve L. Taylor$^3$, Eric J. Terrill$^3$,{\\} Tao Xing$^4$, Shanti Bhushan$^4$, Thomas T. O'Shea$^1$, and {\\} Douglas G. Dommermuth$^1$}

\affiliation{\small $^1$Science Applications International Corporation, USA
\\$^2$Naval Surface Warfare Center - Carderock, USA
\\$^3$Scripps Institution of Oceanography, USA
\\$^4$University of Iowa, USA}

\maketitle

%
%

\section{ABSTRACT}

Full-scale experimental measurements and numerical predictions of the wave-elevation topology behind a transom-sterned vessel, the R/V Athena I, are compared and assessed in this paper. The mean height, surface roughness (RMS), and spectra of the breaking stern-waves were measured in-situ by a LIDAR sensor over a range of ship speeds covering both wet- and dry-transom operating conditions.  Numerical predictions for this data set from two Office of Naval Research (ONR) supported naval-design codes, NFA and CFDship-Iowa-V.4, have been performed. Initial comparisons of the LIDAR data to the numerical predictions at 5.4 m/s (10.5 kts), a wet-transom condition, are presented.  This work represents an ongoing effort on behalf of the ONR Ship Wave Breaking and Bubble Wake program, to assess, validate, and improve the capability of Computational Fluid Dynamics (CFD) to predict full-scale ship-generated wave fields.

%
%

\section{INTRODUCTION}

Full-scale in-situ measurements of the free-surface elevation in the transom region of the R/V Athena I were made in early June of 2005 by research groups from the Marine Physical Laboratory, Scripps Institution of Oceanography-UCSD and the Naval Surface Warfare Center, Carderock Division \cite{Fu06a}. The objective of the experiment was to obtain full-scale qualitative and quantitative breaking transom-stern wave-field data from a naval combatant-like hull form for use in subsequent CFD code development and validation. Although high-resolution measurements of full-scale bow waves \cite**{Fu04} and low-resolution measurements of transom breaking waves have been made in the past \cite**{Fu06b}, this was the first set of high resolution measurements of the breaking transom wave of a full-scale ship.  This paper presents the first comparison of CFD to the transom-stern wave data collected during this experiment.

\begin{figure}
\centering
\includegraphics[width=.9\columnwidth]{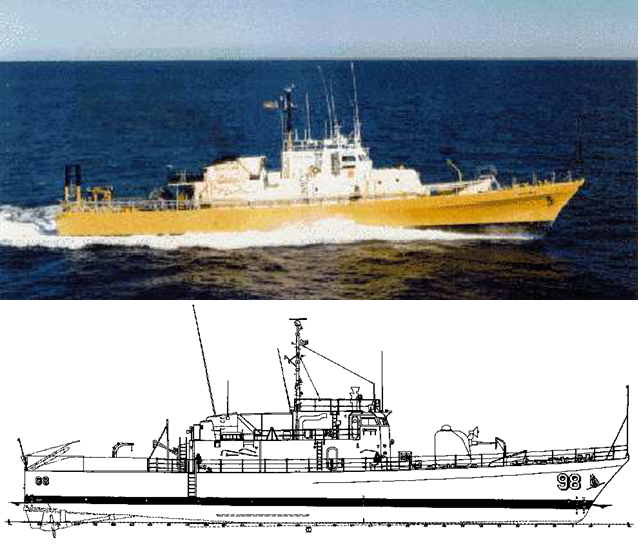}
\caption{\label{fig:Athena}The R/V Athena}
\end{figure}

The vessel used in the experiment, the R/V Athena I, is a converted PG-84 class patrol boat built in 1969 and converted to a research vessel in 1976.  Her hull construction is aluminum with a fiberglass superstructure. The transom-stern of the Athena was used as a source of breaking stern waves for this study.  The Athena's stern has a submergence similar to that of a full-scale DDG-51 class destroyer, making her a relevant selection for transom-stern wave studies.  A picture of the R/V Athena is shown in Figure \ref{fig:Athena}, and the vessel particulars are outlined in Table \ref{tab:details}.

%
%

Measurements of the free surface in the transom region of the R/V Athena were made at various speeds. Table \ref{tab:conditions} lists
these test conditions.  The test set was designed to enable the examination of the transom-stern wake from fully wet to fully dry.  During the experiment the transom was observed to transition from wet to dry transom between 8 to 9 m/s (15 to 17 knots).  This paper will focus on the wetted-transom condition of 5.4 m/s (10.5 knots).

%
%

\begin{table} [h]
\caption {\label{tab:details}R/V Athena Details}
\begin{center}
\begin{tabular}{|c|c|}
\hline \hline
Length Overall  &   50.3 m (165 ft)     \\
\hline
Waterline Length &  46.94 m (154 ft)   \\
\hline
Extreme Beam    &   7.3 m (24 ft)  \\
\hline
Draft           &   3.2 m (10.5 ft) \\
\hline
Propulsion      &   Twin screw \\
\hline
Speed           &   6.2 m/s (12 kts) (diesel) \\
                &   18 m/s (35 kts) (turbine) \\
\hline \hline
\end{tabular}
\end{center}
\end{table}

%
%

\begin{table}
\caption {\label{tab:conditions}Experimental Test Conditions}
\begin{center}
\begin{tabular}{|c|c|c|c|} \hline \hline
Ship Speed & Ship Speed & $F_{r_{L}}$ & $F_{r_{transom}}$ \\
(m/s) & (kts) &   & \\
\hline \hline
3.1 & 6	& 	0.14 & 1.15 \\
\hline
4.6 & 9	& 	0.21 & 1.72 \\
\hline
\textbf{5.4}	 & \textbf{10.5}&	 \textbf{0.25} & \textbf{1.99}\\
\hline
9.3	& 18	&   0.43 & 3.00\\
\hline
13.4	 & 26	&  0.62  & 3.94\\
\hline \hline
\end{tabular}
\end{center}
\end{table}

The sensors deployed for this experiment included the Quantitative Visualization (QViz) system \cite{Furey02}, and a Light Detection and Ranging (LIDAR) sensor \cite**{Fu06b}. Only measurements from the LIDAR will be compared to numerical predictions in this paper.  The LIDAR technique allows scanning of the free-surface at rates rapid enough to resolve the inherent unsteadiness of the breaking transom waves.  The collection of the LIDAR data and its data reduction and analysis are included in the Field Experiment section of this paper.

The full-scale LIDAR data is compared to predictions from two CFD codes currently under development by ONR, Numerical Flow Analysis (NFA) and CFDship-Iowa-V.4. Numerical Flow Analysis is a Cartesian grid formulation of the Navier-Stokes equations utilizing a cut-cell technique to impose the hull boundary conditions. A description of the code and its current capabilities can be found in \citeasnoun{Dommermuth07}. CFDship-Iowa-V.4 is an unsteady Reynolds-Averaged Navier-Stokes (URANS)/detached eddy simulation (DES) code that uses a single-phase level-set method, advanced iterative solvers, conservative formulations, and the dynamic overset grid approach for free-surface flows \cite{Bhushan07}.  The two CFD techniques are compared in separate but complimentary sections in the Numerical Predictions portion of this paper.

%
%

\section{FIELD EXPERIMENT} \label{sec:experiment}

\begin{figure}[h!bt]
\centering
\includegraphics[width=1.0\columnwidth]{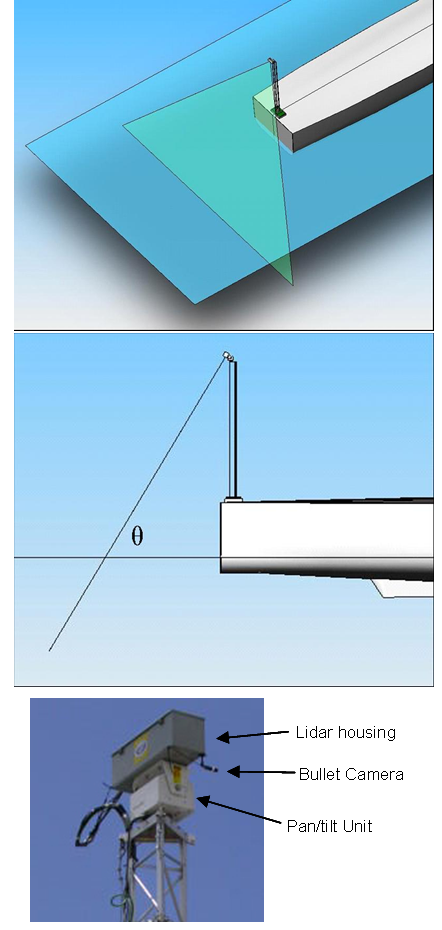}
\caption{\label{Exp_Fig_1}The Scripps LIDAR System mounted to the transom of the R/V Athena I.}
\end{figure}

An at-sea LIDAR-based wave measurement system, developed and deployed by the Scripps Institution of Oceanography, was used to measure the time-varying transom wake at multiple distances aft of the vessel.  The scanning LIDAR system consists of a 2D scanning LIDAR unit (2.5 cm range resolution), a pan/tilt unit to control the measurement location to within {$\pm$}{$0.25^{\circ}$} in elevation and azimuth, a bore-sighted camera for visualizing the field scanned by the LIDAR, a GPS for determining vessel speed, and a time-synchronized 6-degree of freedom motion sensor.  The other transom measurement systems deployed during the field tests have been described by \citeasnoun{Fu06a}.  The LIDAR system was mounted on top of a tower approximately 12 m above the free surface and 0.58 m port of the ship centerline.  A drawing of the Scanning LIDAR System onboard the Athena is shown in Figure \ref{Exp_Fig_1}.

%
%

The LIDAR enploys a Class I 900 nm wavelength Laser (Riegl LMS-Q140i-80), with an 80-degree maximum swath width, and a 20 Hz sampling rate.  The pan/tilt unit (QuickSet QPT-90IC) has an integrated controller which enabled remote control of the elevation and azimuth angles of the LIDAR.  An inertial motion unit (Xbow motion sensor) was packaged with the LIDAR head unit at the top of the tower and sealed inside of a weatherproof housing (Figure \ref{Exp_Fig_1}).  The unit was placed at the top of the tower to directly measure any potential tower vibration that could not be recorded by the principal ship-motion sensor, a higher grade GPS-6 DOF unit (CodaOctopus F180), affixed to the deck of the vessel.

The LIDAR system was operated in two modes:  1) sweeping mode, in which the LIDAR was continuously swept along ship centerline from the transom to the aft-horizon, using the pan/tilt, effectively scanning a swath of the wake up to 12 m aft of the vessel; or 2) fixed angle mode, where the LIDAR scanned the free surface at a constant elevation and azimuth relative to the ship.  The programmable pan/tilt unit allowed data to be sequentially captured, typically gathering 60 seconds of data at a number of preset elevation angles.   The flexibility of these deployment schemes allowed the interrogation of different regions within the wake.  Positioned at almost 12 m above the mean waterline, the LIDAR's effective swath was approximately 20 m wide.  Theoretically, the field-of-view should increase further aft, but data returns decreased as the angle-of-incidence with the water surface increased.  Thus 20 m was the effective maximum swath that the LIDAR could measure for the tower height used in these studies.

Mean wake profiles were obtained by the LIDAR system at 3.1, 4.6, 5.4, 9.3, and 13.4 m/s (6, 9, 10.5, 18, and 26 kts).  In this paper the 5.4 m/s (10.5 kt) dataset is presented and discussed.

\subsection{Data Processing and Analysis}

The post-processing scheme transforming the raw LIDAR data to wave height involved 1) conversion of the raw data to earth-based coordinates moving with the vessel, 2) correcting the free surface to mean sea level, and 3) compositing the surface at various locations aft of the vessel into a time-averaged 3D map.

The scanning LIDAR outputs range and angle of the rotating mirror.  These data were converted to ship Cartesian coordinates by using the elevation and azimuth angles of the pan and tilt.  The vertical axes of the LIDAR and the ship-fixed coordinate system were aligned, with {$x$} positive aft, {$z$} positive downward, and the origin at the LIDAR.  Measurements from the F180 (a combined GPS/motion instrument that, in addition to GPS information, outputs acceleration in three axes, pitch, roll, yaw, and heave in real-time) was used to correct the data to a ground-fixed system.  Real-time heave was calculated from the acceleration data and simultaneously corrected for drift using a Kalman filter.

The registration of time within the LIDAR system was required for accurate motion correction of the data.  The Xbow Dynamics Motion Unit (DMU) and LIDAR were synchronized by an external trigger which provided an accurate time registration in the data acquisition coincident with the reset of an internal clock within the LIDAR.  The clock reset trigger was typically used at least twice during an 11-minute data collection run.  Upon examining the timing synchronization between the LIDAR and DMU clocks, it was found that the LIDAR clock drifted less than 50 msec in any 11-minute period, i.e. the drift never exceeded the temporal spacing between LIDAR measurements.  Due to the nature of the scanning mechanism, there is an along-track translation between the first and last pixels of a single line scan.  The time between the first and last pixel was ~1.47 msec; for a maximum ship speed of 15 m/s, the longitudinal transit of a scan due to ship translation was 2.21 cm.  The nominal footprint of the laser on the ocean surface varies with distance aft and distance away from the center scan, but it was generally 3-4 cm in diameter.  Due to the small translation distance relative to the footprint size, each line scan was treated in post processing as an instantaneous point measurement.

During processing of data obtained when the wake is scanned with the pan/tilt unit, careful attention is require to register the time and position of the LIDAR head angle.  The approach used relied upon the DMU's $x$-axis (which was parallel to the LIDAR's centerline measurement) acceleration, whose second derivative clearly shows the start/stop times of elevation changes.  In fixed-angle mode, a relatively constant $x$-acceleration was assumed to signify a constant elevation angle.  In the sweeping mode, a linear rate of change in elevation was assumed, which was a reasonable assumption for the angles used in this data collection.

The DMU/LIDAR data also required time synchronized with an external GPS timestamp.  The approach was to cross-correlate $x$-axis data from the two motion packages to obtain the constant time-offset (typically 1 second or less) between the LIDAR data and the F180/ship motion data.  Once the timing of the LIDAR measurement system with the F180 GPS time reference was reconciled, now to an accuracy of O(0.01) second, the raw LIDAR measurements could be converted to an earth-based coordinate system using the F180 pitch and roll angles.

In the second post-processing step, the measured free surface was corrected to Mean Sea Level (MSL).  MSL would ideally be measured when the boat was stationary, but data was only collected at non-zero ship velocities; so the average waterline measured approximately 2 m aft of the transom at a ship speed of 1.0 m/s (2.0 kts) was used as the MSL vertical datum.  Data was collected for MSL over five minutes.  In the final post-processing step, a 3D map of the mean surface wave field was generated from elevation measurements at a variety of LIDAR angles.  A rectangular grid was created and the field data were placed onto the grid and averaged within each grid cell.  The size of each grid cell was 0.1 m wide by 1 m long.  Empty grid cells were linearly interpolated, and a 3-point median filter was applied along the wake's width to create the final images.
	
\subsection{LIDAR Results}

\begin{figure}[h]
\centering
\includegraphics[width=1.0\columnwidth]{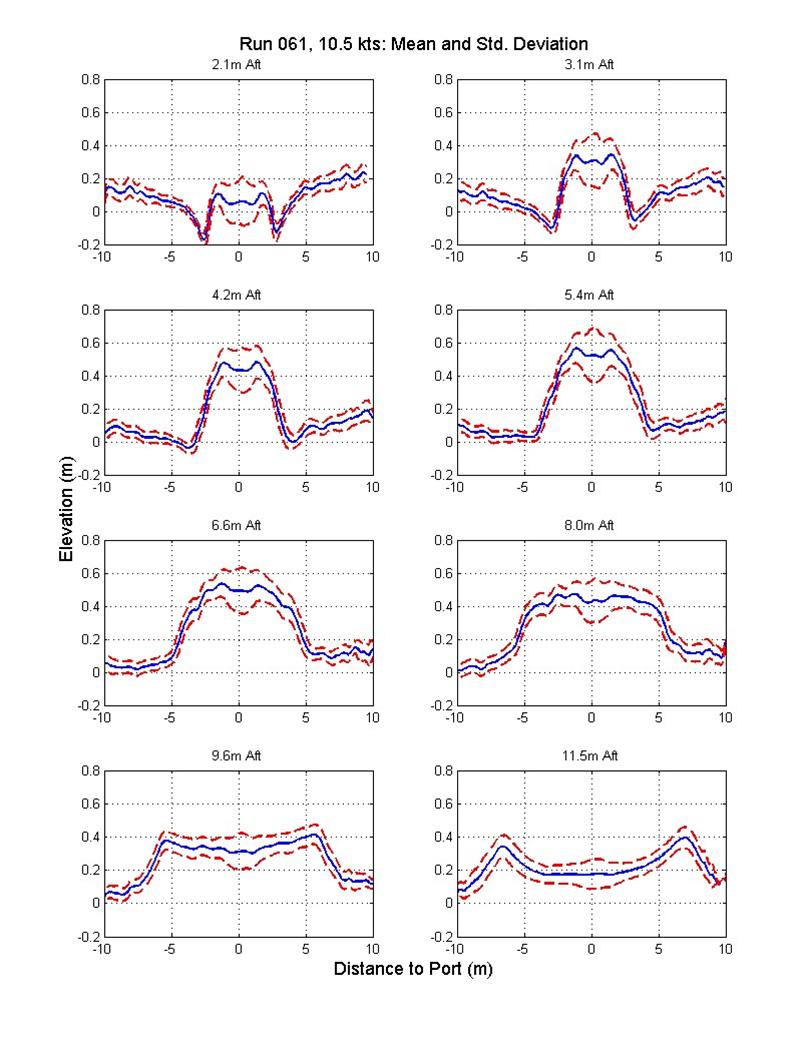}
\caption{\label{Exp_Fig_2}Mean (blue) and standard deviation (red) of the wake surface at various aft locations.}
\end{figure}

Figure \ref{Exp_Fig_2} shows mean wake profiles at locations ranging from 2.1 m aft to 11.5 m aft of the transom.  Standard deviation is also included as the red dotted line.  For this particular data set, the LIDAR system was set to collect data in time segments at fixed angles.  Elevations ranged between -80∞ to -45∞ in 5-degree increments.  The aft location given was the mean location of the free-surface interrogated by the LIDAR at each angle.  Since the along-scan resolution varies with time and elevation angle, the LIDAR data were gridded to 0.1 m resolution in the port ($y$-axis) and the mean and standard deviation of elevation were calculated for each grid cell.  Only locations within {$\pm$}10 m from the centerline are shown because the data drop-out rate was much higher outside this region leads to unreliable measurement statistics

%
%

In general, wake elevations are higher towards ship centerline and decrease away from the center.  The local maximum to either side of the centerline is the wake's "shoulder."  Notice a slight depression in height between the two shoulders.  Just outside the shoulder, the wake elevation decreases sharply, and continuing outward from the center, the surface remains fairly constant (as at 6.6 and 8.0 m aft) or rises gradually (3.1, 4.2, 5.4 m aft).  Standard deviation is fairly constant across this region of the wake except approximately {$\pm$}2 m around the center.  A minimum in standard deviation is present near the trough area just outside of the wake shoulders.  The distance between the shoulders increases with aft distance, which is characteristic of a ship's V-shaped wake.

%
%
\begin{figure}[h!bt]
\centering
\includegraphics[width=1.0\columnwidth]{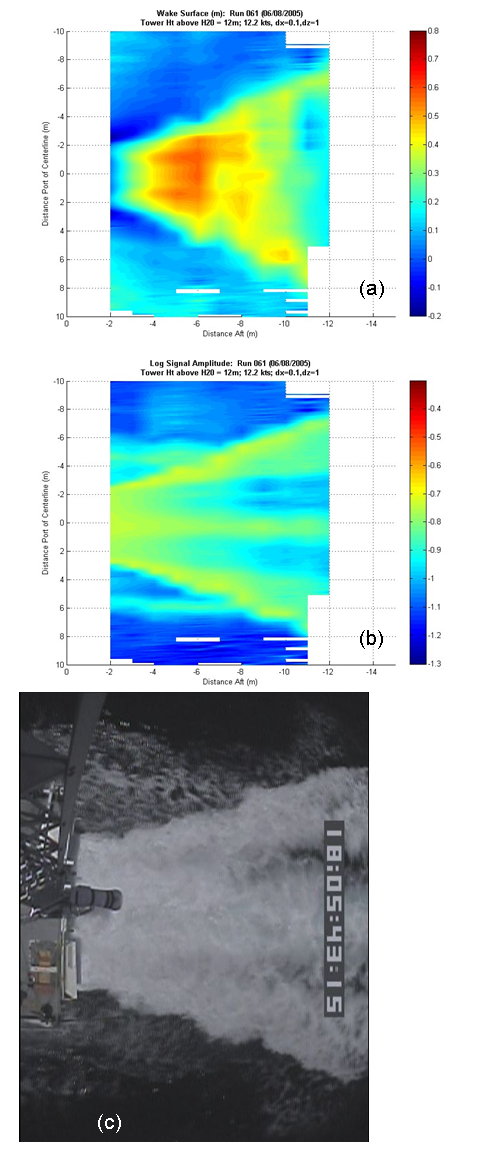}
\caption{\label{Exp_Fig_3}(a) Mean wake surface, (b) logarithmic signal return amplitude, and (c) photograph at 5.4 m/s (10.5 kts). }
\end{figure}

%
%
\begin{figure} [h!bt]
\centering
\includegraphics[width=1.0\columnwidth]{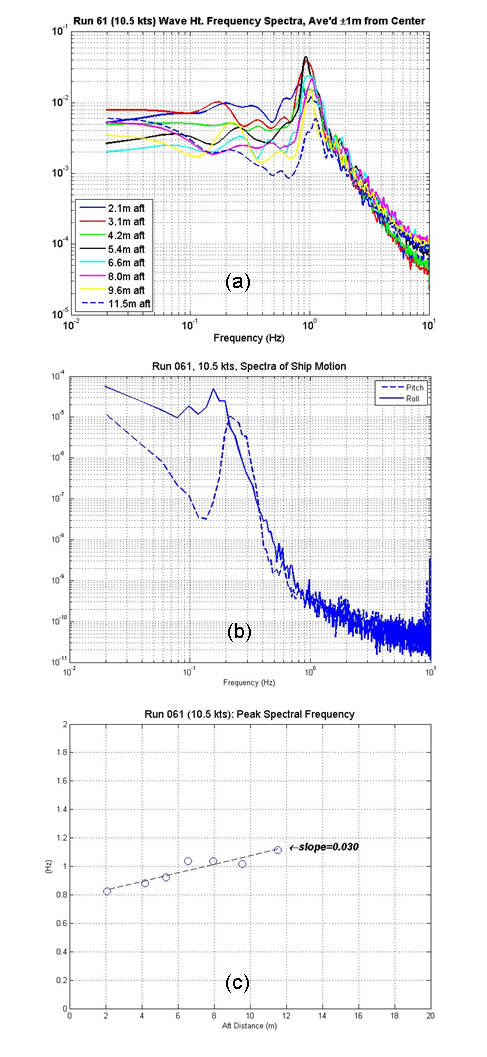}
\caption{\label{Exp_Fig_4}(a) Frequency spectra of the wake elevation at various aft locations.  Spectra were averaged spatially between 1 m port and 1 m starboard of ship centerline. (b) Pitch, roll spectra. (c) Peak of the frequency spectrum vs. aft distance.}
\end{figure}

The elevation data at all tilt angles were inserted into a regularly-spaced 2D grid to composite the free surface.  The V-shape of the wake is well-defined in the data (Figure \ref{Exp_Fig_3}a) and in the photograph from the LIDAR system's camera (Figure \ref{Exp_Fig_3}c).  The trough outside the shoulders and the slight centerline depression can also be observed.  Figure \ref{Exp_Fig_3}b is a reconstruction of the average normalized signal return (on logarithmic scale) to the LIDAR.  This figure is instructive because the brightness of the signal return is dependent on the optical scattering efficiency of the free surface, which is dependent on the concentrations of bubbles and foam at the surface.  The data from this active optical technique shows the bubbles are mostly found at the shoulders and along the centerline, consistent with the passive imaging provided by the black/white photograph.  Additionally, within 7 m aft of the transom, a streak of bubbles can be seen at 5 - 6 m port and starboard of the centerline.  This was also observed in the video footage (Figure \ref{Exp_Fig_3}c).  These bubbles were created either at the bow or along the side hull of the vessel and were being swept aft to the transom region.

With the time-series of surface elevation having been obtained at multiple port locations, the spectral content across the wake was computed, and it was found that the most energetic portion of the wake was in a region of {$\pm$}1 m along the centerline of the wake (an observation consistent with the previously discussed standard deviations shown in Figure \ref{Exp_Fig_2}).  Frequency spectra for this region at multiple distances aft of the vessel are shown in Figure \ref{Exp_Fig_4}a.  The peaks in the spectra are close to 1 Hz.   A slight linear relationship is observed between distance aft and peak frequency (Figure \ref{Exp_Fig_4}c).  The absence of a corresponding peak at 1 Hz in the pitch/roll spectra, Figure \ref{Exp_Fig_4}b, suggests that this peak is a characteristic of the inherent unsteadiness of the wake, and not due to ship motions.

\subsection{Field Data Discussion}

\begin{figure} [hbt]
\centering
\includegraphics[width=1.\columnwidth]{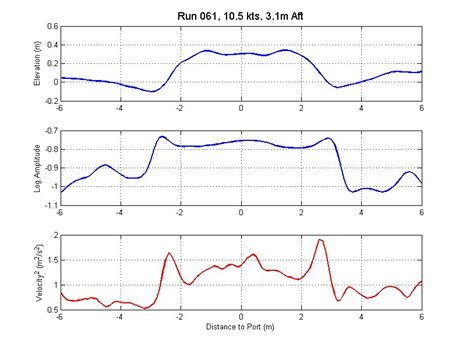}
\caption{\label{Exp_Fig_5}Mean elevation, logarithmic normalized signal amplitude, and square velocity at 3.1 m aft of the transom, 5.4 m/s (10.5 kts).}
\end{figure}

\begin{figure} [hbt]
\centering
\includegraphics[width=1.\columnwidth]{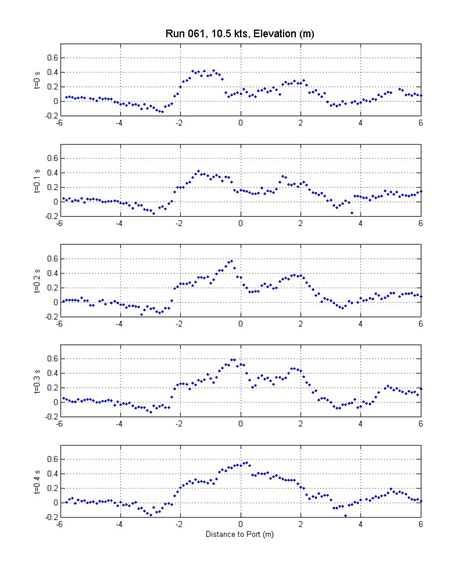}
\caption{\label{Exp_Fig_6}Instantaneous wake elevations every 0.1 sec.  Two waves from the shoulders are traveling in opposite directions.}
\end{figure}

Figure \ref{Exp_Fig_2} illustrates that the central region of the wake's surface is the most variable, while the regions outside the shoulders have relatively low variability.  This variability is found to result from a narrowband wave spectrum (Figure 5) in the frequency range between 0.8 - 1.1 Hz. Time series analysis of the LIDAR data, and qualitative observations from bore-sighted video camera indicate that breaking shoulder waves traveling to the vessel centerline collide.  When the opposing waves meet in the center, a peak in the surface profile forms (Figure \ref{Exp_Fig_6}).  Subsequently there is a depression in the center of the profile as the water falls away from the center.  Thus a standing wave is observed that occurs between the wake's shoulders with a fundamental frequency of 0.8 - 1.1 Hz.  Analysis of the square of the vertical velocity along the cross-track of the wake (computed using the time derivative of the elevation measurements) indicates there is a spike in kinetic energy just outside of the wake shoulders and near the centerline (Figure \ref{Exp_Fig_5})where breaking is most active.  Consistent with observations of a standing wave, it appears that the nodes of the standing wave are approximately {$\pm$}3 m from center, where the mean elevation and kinetic energy are at a minimum.  Moreover, the sloping edge of the shoulders at approximately {$\pm$}2.5 m from center are areas of active breaking, as indicated by the kinetic energy spikes.  Active breaking also occurs near the centerline as the two opposing waves from the shoulder collide.
%
%

%
%

%
%

\section{NUMERICAL PREDICTONS} \label{sec:numerics}

Initial comparisons of the experimental LIDAR measurements are made to predictions from NFA and CFDship-Iowa-V.4 in the following sub-sections.

In the previous section, a significant amount of dynamic heave, roll, and pitch was removed from the LIDAR data to register the data to MSL.  The fact that such a small final standard deviation was achieved suggests this registration was applied successfully (see Figure \ref{Exp_Fig_2}). However, the data in that figure displays both a higher than expected (or predicted) mean height in addition to a clear roll dependence. It was therefore deemed necessary to remove these last biases before making the quantitative comparisons.

In the following sections, the  LIDAR mean-height data has been de-trended in the following manner:  The mean elevation data has been reduced in height by 0.08 m, and a roll of 0.28 degrees has been removed.  The resulting heave correction of -0.08 m is of the same order as the dynamic heave difference measured at model-scale from the 0 m/s to 5.4 m/s (10.5 kts) operation (-0.05 m to -0.075 m).  Although the F180 could measure ship response to most ocean waves, it is surmised that the heave due to steady ship dynamics may have been incorrectly interpreted as sensor drift and therefore not taken into account.  The 0.28 degree tilt could also have arisen from a small alignment difference between the sensor with the boat hull.

    %
    %

\subsection{NFA Predictions} \label{subsec:NFA}

The objective of the numerical predictions is to assess the capability of NFA to predict unsteady transom flows of model-scale and full-scale ships.

A description of the NFA and its current capabilities can be found in \citeasnoun{Dommermuth07}.  NFA solves the Navier-Stokes equations utilizing a cartesian-grid formulation to model the free surface and a ship hull.  An interface capturing technique is used to resolve the free surface.  The interface capturing of the free surface uses a second-order accurate, volume-of-fluid technique.  A cut-cell method is used to enforce free-slip boundary conditions on the hull.  The free-slip boundary condition does not resolve the ship's boundary layer.   Away from the hull, NFA uses a Smargorinsky turbulence model based on a large-eddy formulation.  A surface representation of the ship hull is all that's required as input.  Relative to methods that use a body-fitted grid, the potential advantages of NFA's approach are significantly simplified gridding requirements and greatly improved numerical stability due to the highly structured grid.

As an initial step in evaluating the capability of NFA to model the wake of the full-scale Athena, a grid-density convergence study was performed.  Two simulations have been performed with medium and high grid densities for the Athena moving with constant forward speed at 5.4 m/s (10.5 kts) ($F_r=0.2518$), and one simulation has been performed with medium resolution for 9.3 m/s (18 kts) ($F_r=0.4316$).     The medium-resolution simulations used 1024x192x256=50,331,648 grid points, and the high-resolution simulation used 1280x384x384 = 188,743,680 grid points.     The length scales were normalized by the ship length ($L=46.939$ m), and the velocity scales were normalized by the ship speed (U=5.4 m/s (10.5 kts) \& 9.3 m/s (18 kts)).   Grid stretching was used along the cartesian axes.   Grid points were clustered near the stern region, along the sides of the ship,  and the mean free-surface.     The minimum grid spacing, {$\Delta$}, for the high-resolution simulation  ($\Delta=0.000499$) was about half the medium simulation ($\Delta=0.00109$).   For the high-resolution simulation, the grid spacing near the edges of the domain was ($\Delta=0.00831$), which due to more stretching was slightly higher than the medium-resolution simulation ($\Delta=0.00656$).   For all simulations, the length, width, water depth, and air height of the computational domains were respectively 3.0, 1.0, 1.0, and 0.5 ship lengths.  The non-dimensional time steps were $\Delta t=0.00025 \; \& \; 0.000125$, and the numerical simulations run 28,000 and 32,000 time steps for the medium and high-resolution simulations, respectively.  This corresponds to seven ship lengths for the medium-resolution simulations and four ship lengths for the high-resolution simulation.  An adjustment procedure was used to bring the ship up to full speed.  The adjustment period was 0.5.     The results of the medium-resolution simulation were output every 20 time steps. Due to file size consideration, the high-resolution simulation was only output every 100 time steps.  The sampling rates for the medium and high-resolution simulations were respectively  23.0137 Hz and 9.2055 Hz in dimensional units.  The medium and high-resolution simulations respectively used 384 and 720 subdomains.  For each simulation, each subdomain was assigned to a single node on a Cray XT3.  The medium and high-resolution simulations respectively take 46.7 and 122 wall-clock hours, which corresponds to 6.0 and 13.8 cpu seconds per time step.

For this investigation: symmetry boundary conditions were imposed across the centerplane of the ship, appendages and propulsors were not modeled, and the hull was fixed in sinkage and trim.  The sinkage and trim conditions were established based upon high-density Das Boot \cite{Wyatt00} predictions.  At 5.4 m/s (10.5 kts) the sinkage was set at $-9.67^{-4}$ (in non-dimensional units), and trim was set at $0.1031^{\circ}$.  At 9.3 m/s (18 kts), the sinkage was $-3.15^{-3}$ (non-dimensional), and the trim was $0.4510^{\circ}$.

%
%

\begin{figure} [h!]
\begin{center}
\begin{tabular}{rc}
(a) & \vspace{-16pt} \\
& \includegraphics[width=0.8\linewidth]{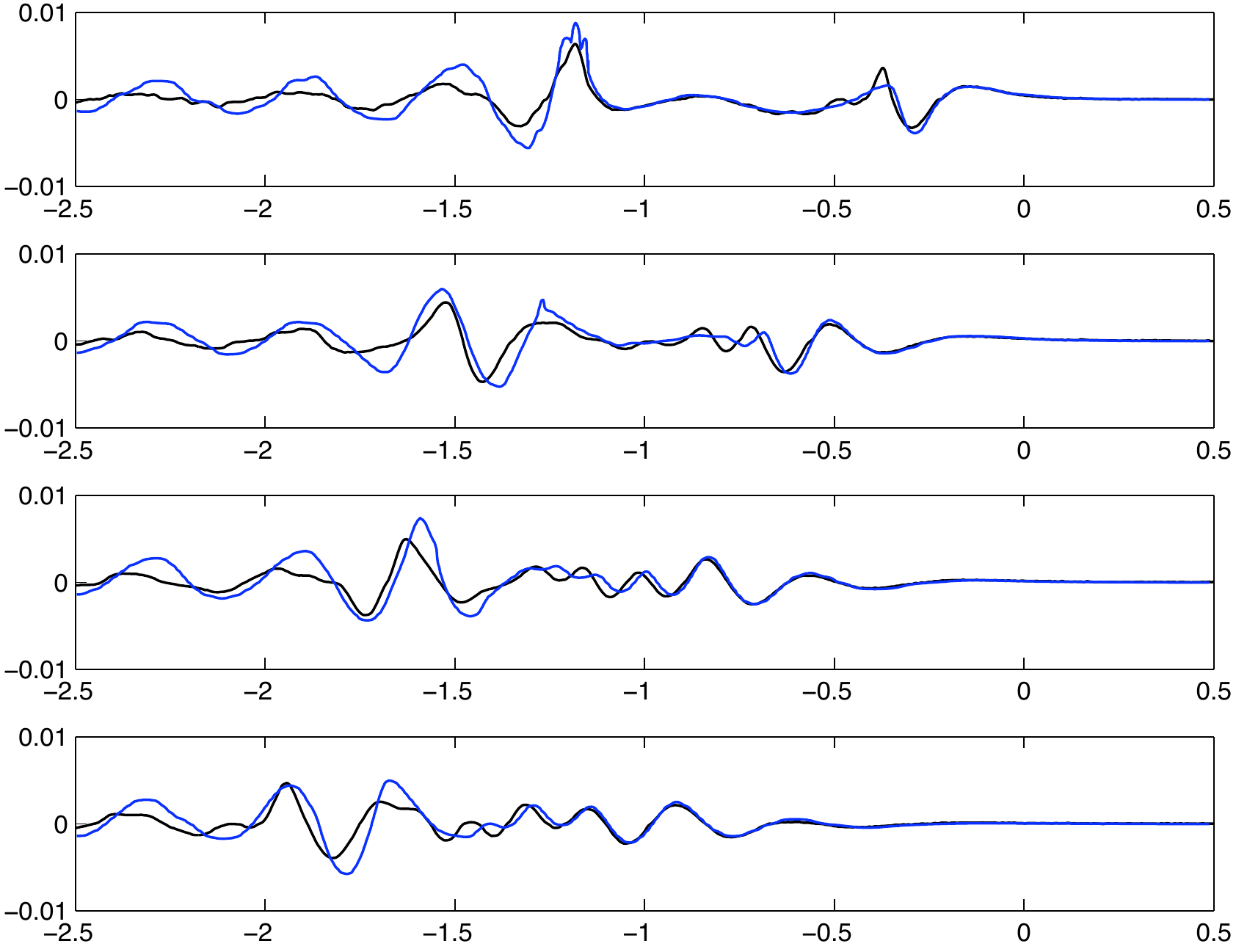} \\ & \\ & \\
(b) & \vspace{-16pt} \\
& \includegraphics[width=0.8\linewidth]{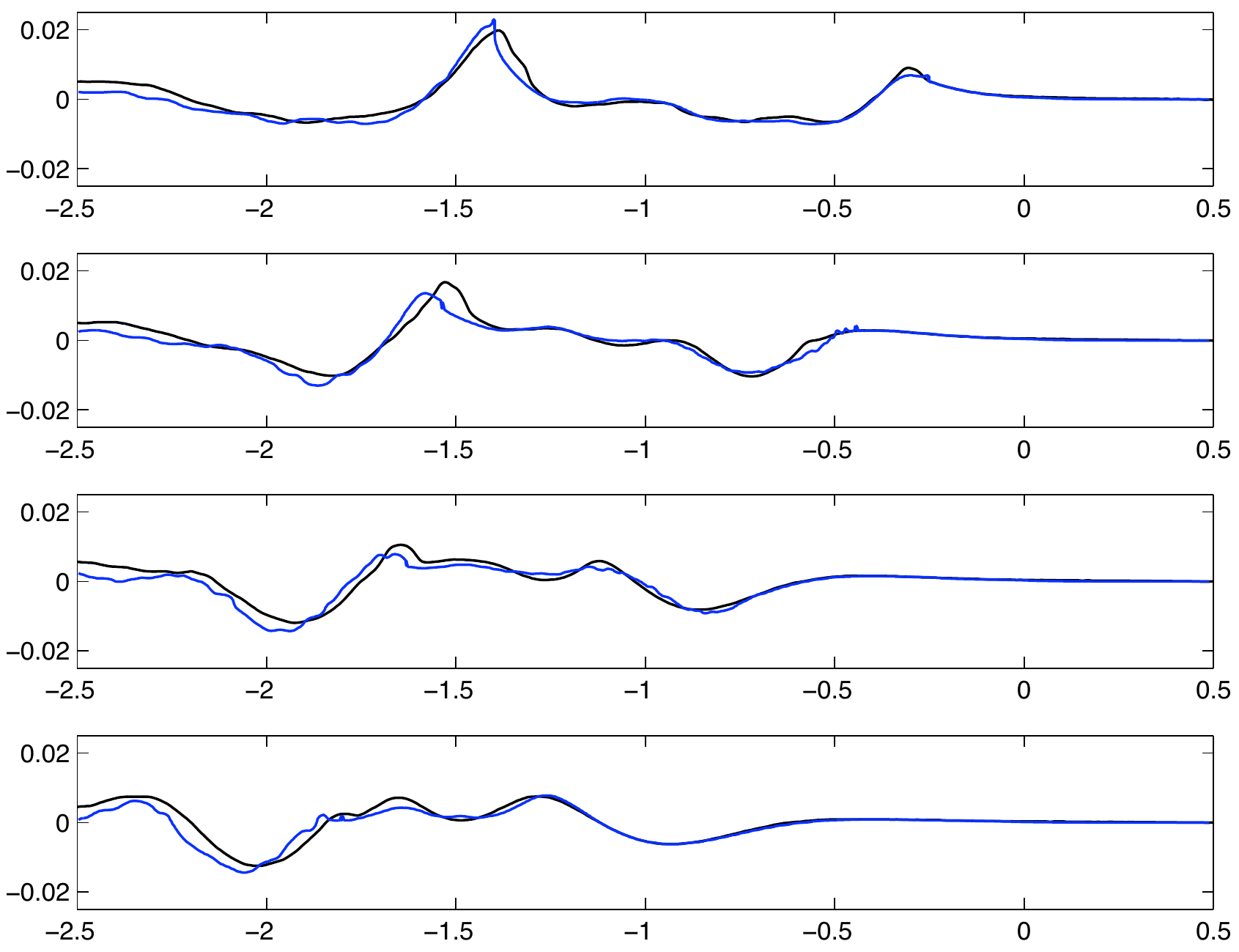}
\end{tabular}
\caption{\label{nfa_cuts} Model 5365 (Athena) wave cuts at four transverse offsets.   NSWCCD measurements and NFA predictions are respectively denoted by black and blue lines.  The fore perpendicular is located at \mbox{x/L=0} and the aft perpendicular is located at \mbox{x/L=-1}.   For each set of plots, the transverse offsets are \mbox{y/L=0.1067}, 0.1861, 0.2482, and 0.31023.  (a) 5.4 m/s (10.5 kts), full-scale equivalent.  (b) 9.3 m/s (18 kts), full-scale equivalent.}
\end{center}
\end{figure}

\subsubsection{Model-Scale Predictions and Comparisons}

\begin{figure} [h!]
\begin{center}
\begin{tabular}{rc}
(a) & \vspace{-16pt} \\
& \includegraphics[width=0.8\linewidth]{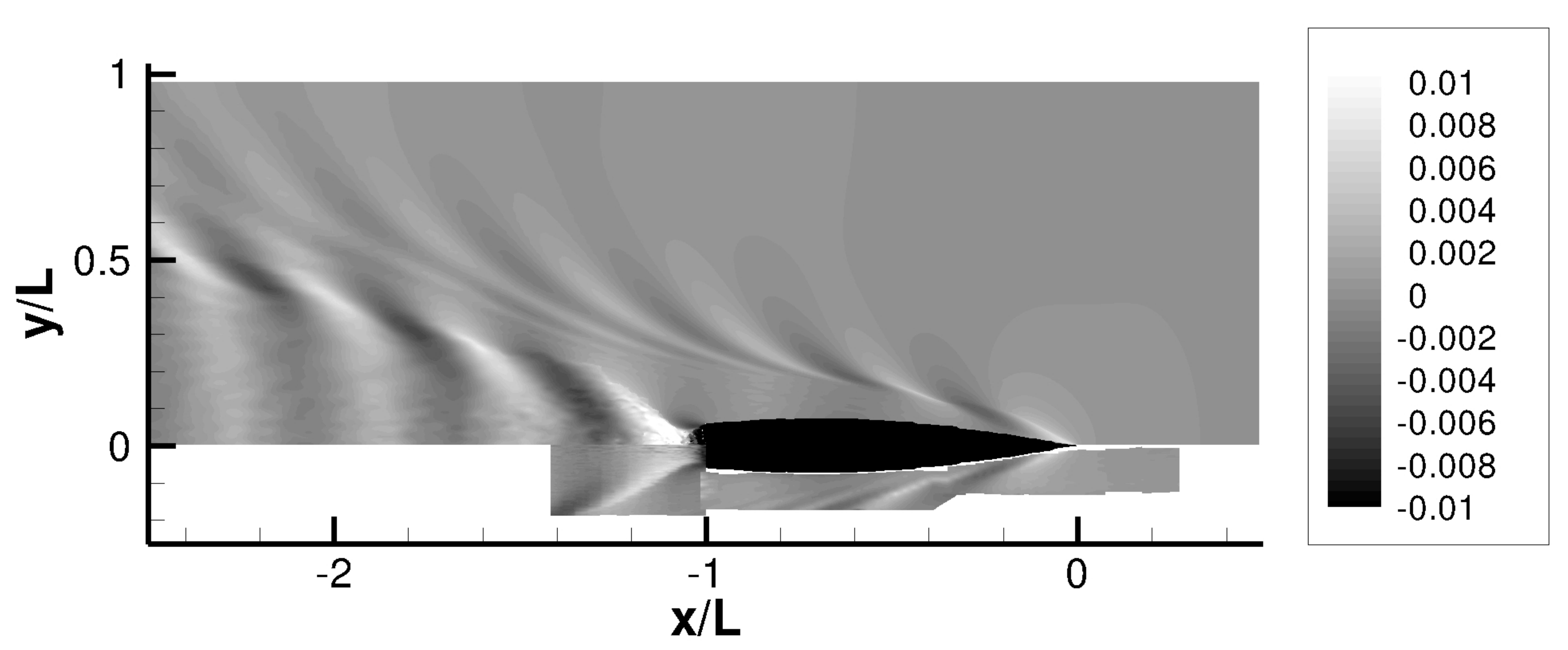} \\
(b) & \vspace{-16pt} \\
& \includegraphics[width=0.8\linewidth]{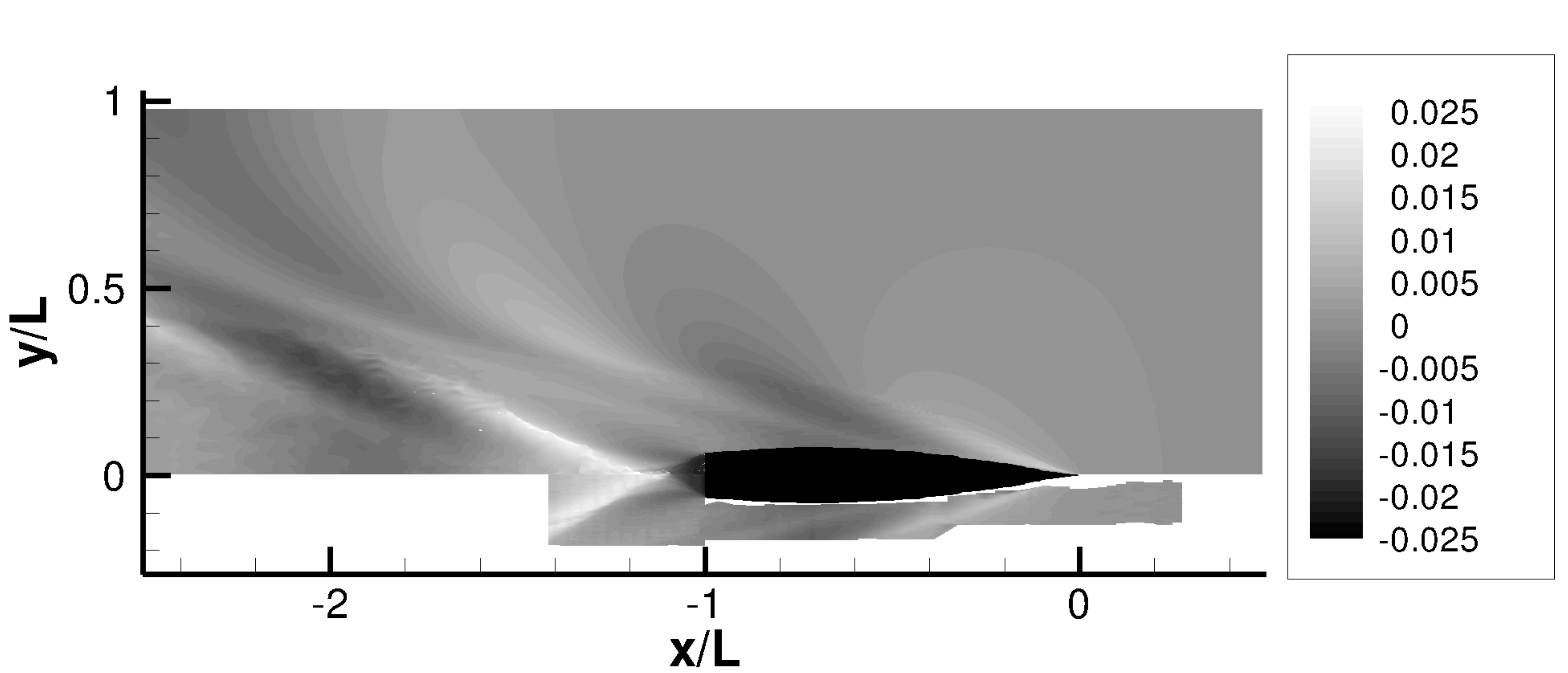}
\end{tabular}
\caption{\label{nfa_contours} Model 5365 (Athena) free-surface contours.  NSWCCD measurements and NFA predictions are respectively plotted in the bottom and top halves of each plot.  (a) 5.4 m/s (10.5 kts), full-scale equivalent.  (b) 9.3 m/s (18 kts), full-scale equivalent.}
\end{center}
\end{figure}

Figure \ref{nfa_cuts} compares NSWCCD model-scale bare-hull measurements and NFA predictions of wave cuts for the 5.4 m/s (10.5 kt) and 9.3 m/s (18 kt) cases.  Results are shown for medium grid resolutions.  The agreement between measurements and predictions is better for the 9.3 m/s case than for the 5.4 m/s case.  Apparently, the dry-transom flow at 9.3 m/s (18 kts) is predicted better by NFA than the wetted-transom flow at 5.4 m/s (10.5 kts).  In fact, for 5.4 m/s (10.5 kts) the transom is dry as predicted by NFA, whereas a wet transom is observed at model-scale.
Figure \ref{nfa_contours} compares NSWCCD bare-hull measurements and NFA predictions of free-surface contours for the two speed cases.   Results are shown for medium grid resolutions.  Once again, the agreement between measurements and predictions is better for the 9.3 m/s (18 kt) case than for the 5.4 m/s (10.5 kt) case.   The contour plots show no evidence of numerical dissipation away from the ship model.

%
%

\subsubsection{Full-Scale Predictions and Comparisons}

\begin{figure*}
\begin{center}
\begin{tabular}{rcrc}
(a) & & (b) & \vspace{-16pt}  \\
& \includegraphics[width=0.3\linewidth]{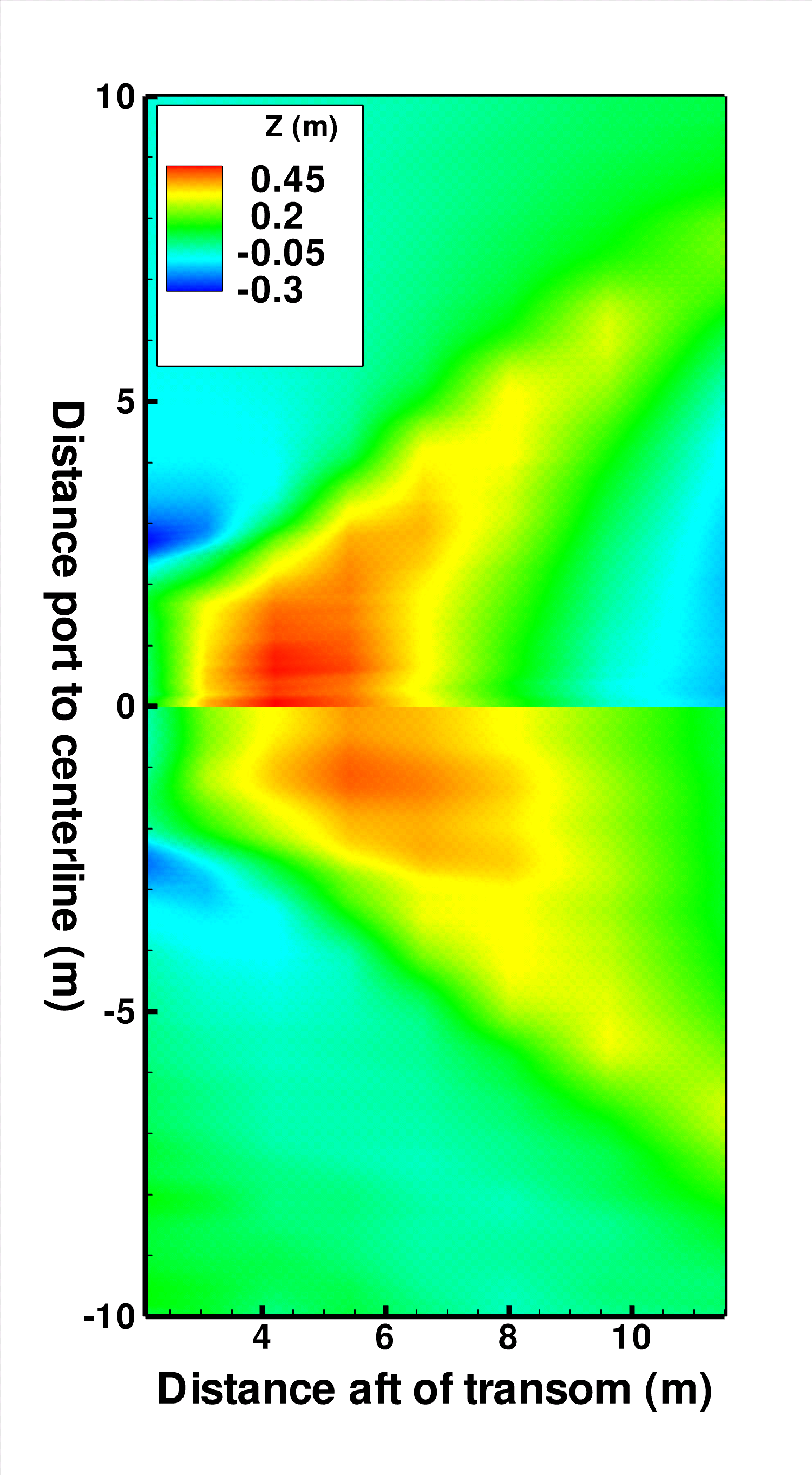}
& & \includegraphics[width=0.3\linewidth]{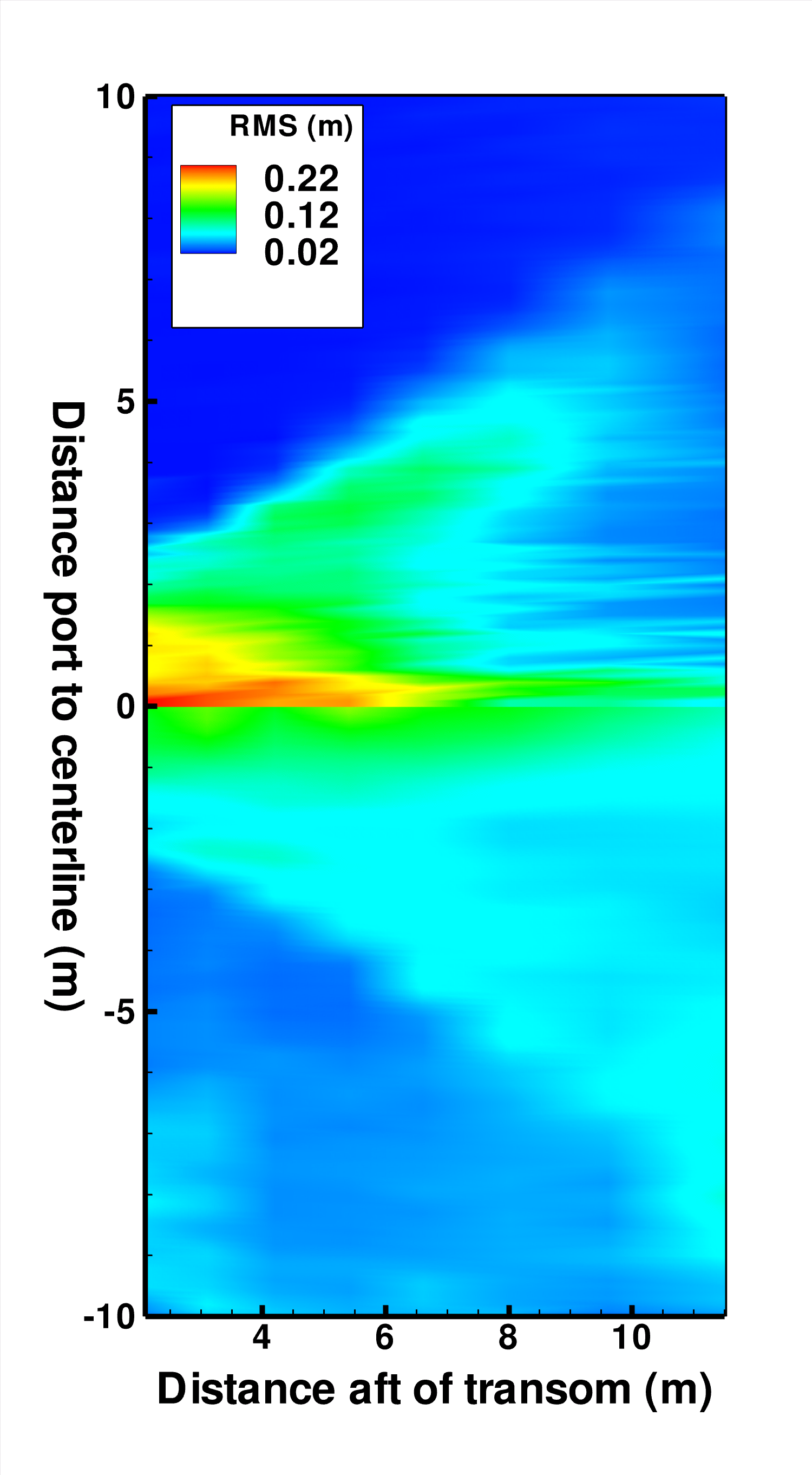} \\
(c) & & (d) & \vspace{-16pt}  \\
& \includegraphics[width=0.3\linewidth]{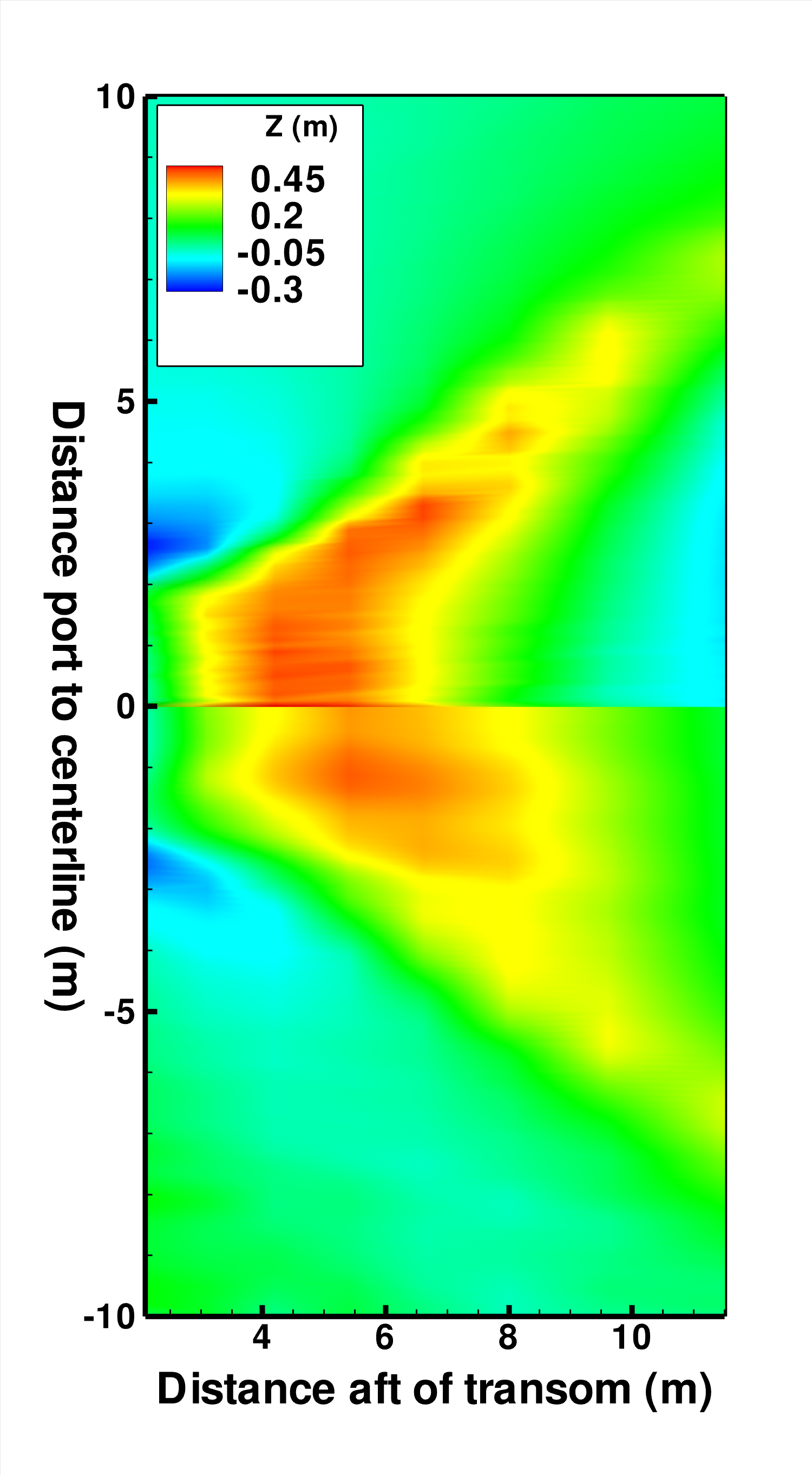}
& & \includegraphics[width=0.3\linewidth]{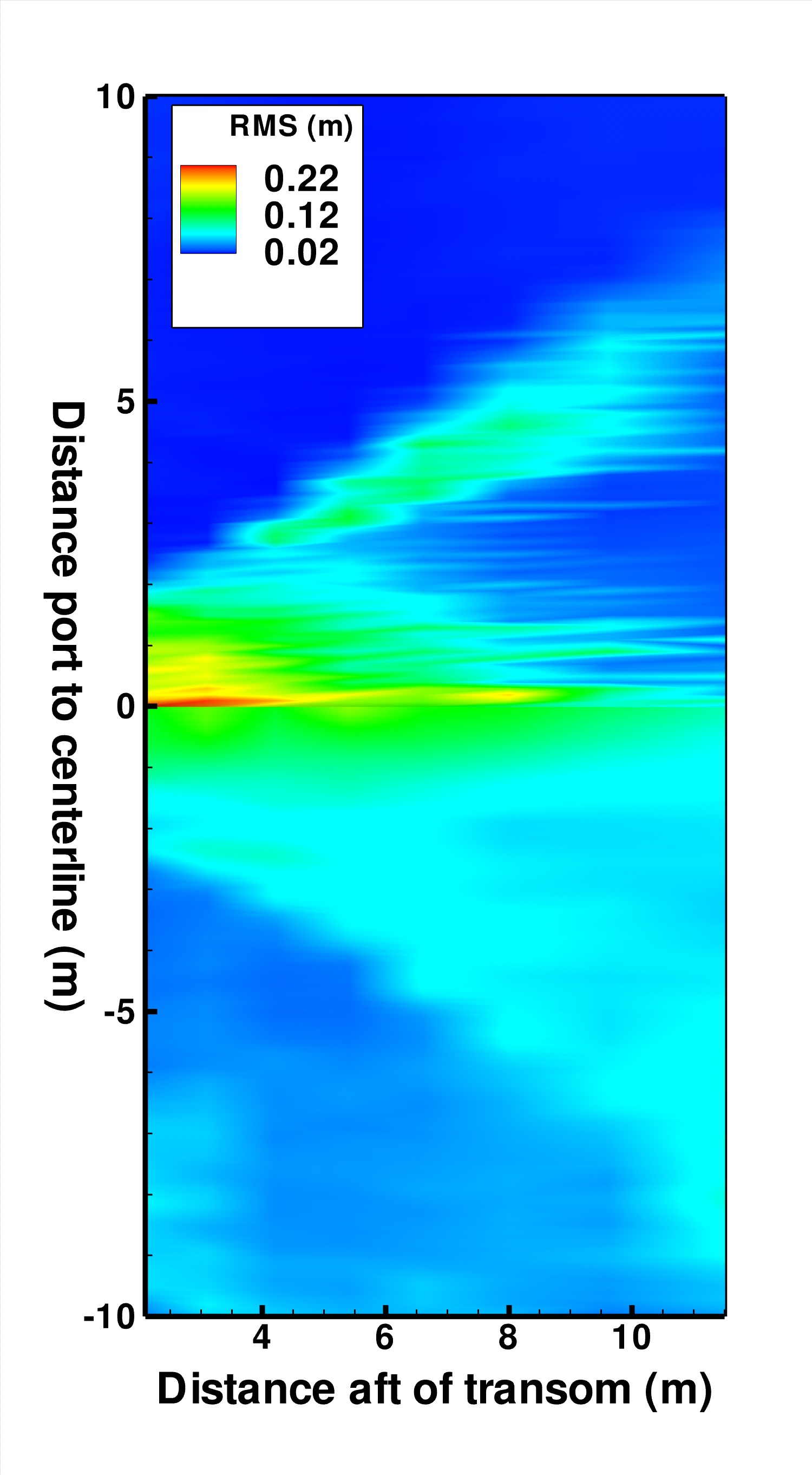}
\end{tabular}
\caption{\label{nfa_lidar} NFA versus experiments in the stern region at 5.4 m/s (10.5 kts).   NFA predictions and LIDAR measurements are respectively plotted in the top and bottom halves of each plot.  (a) Wave height (medium resolution, correlation coefficient=0.81).  (b) RMS height (medium resolution, correlation coefficient=0.88). (c) Wave height (high resolution, correlation coefficient=0.84). (d) RMS height (high resolution, correlation coefficient=0.85). }
\end{center}
\end{figure*}

Figure \ref{nfa_lidar} compares LIDAR measurements and NFA predictions of the free-surface elevation in the stern region for the 5.4 m/s (10.5 kt) case.   Mean and RMS free-surface elevations are shown for medium and high-resolution simulations.   For the mean free-surface elevation, the high-resolution prediction are in slightly better agreement with field measurements than the medium-resolution prediction.   The opposite is true for the RMS fluctuations in the free-surface elevation.   Both the predicted mean and RMS free-surface elevations attenuate more rapidly downstream than the field measurements.  The predicted RMS fluctuations are higher than the measurements along the centerplane.  This is where symmetry has been imposed in the NFA simulations.  We also note that the background RMS fluctuations are much higher in the field measurements than the numerical predictions.  Due to the collection of measurements in an unsteady environment, some background level of RMS surface roughness is to be expected.

%
%

\begin{figure} [h]
\begin{center}
\begin{tabular}{rc}
a) & \vspace{-16pt} \\
& \includegraphics[width=0.85\columnwidth]{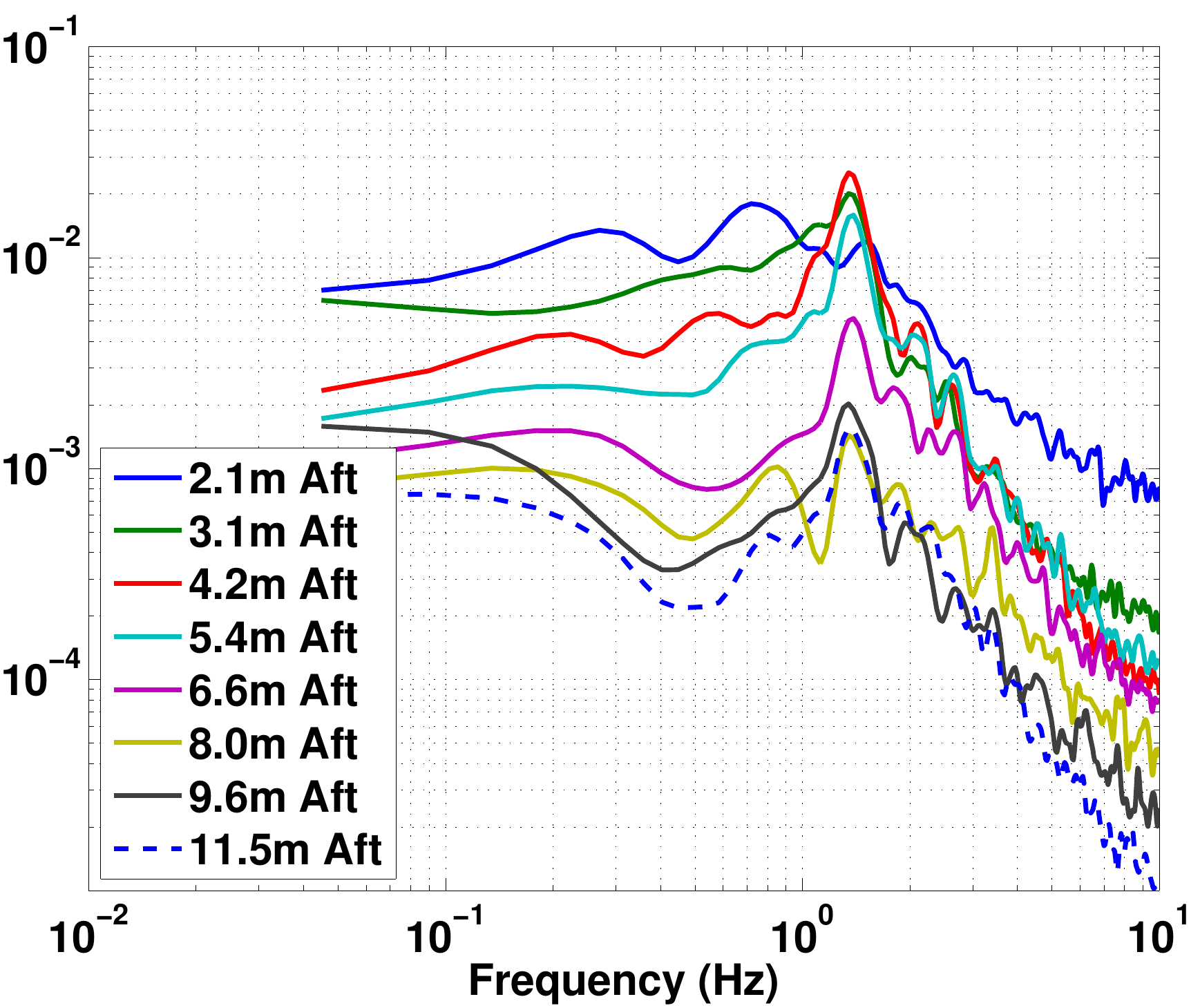} \\
(b) & \vspace{-16pt} \\
& \includegraphics[width=0.85\columnwidth]{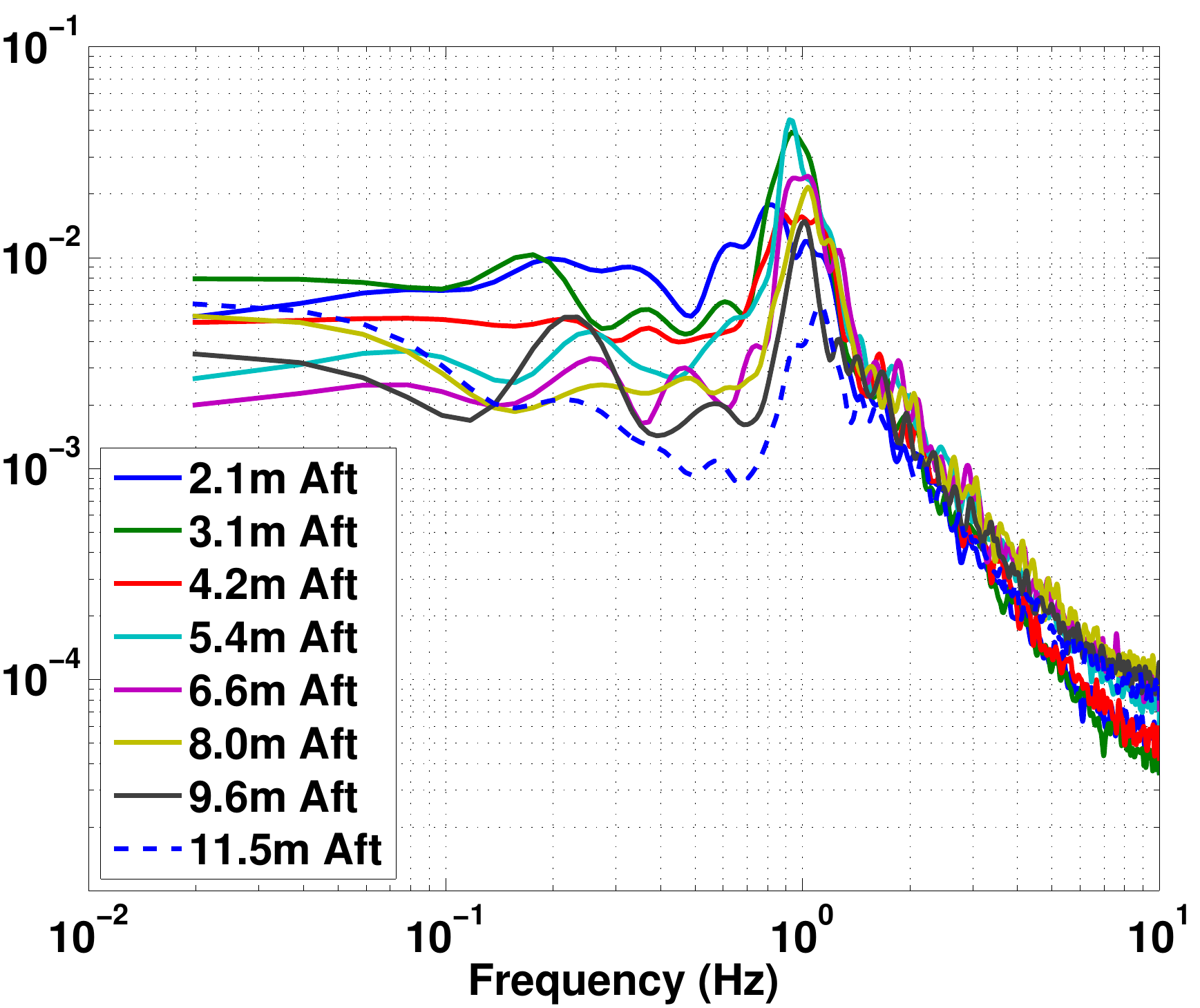}
\end{tabular}
\caption{\label{nfa_lidar_spectra} NFA versus experiments in the stern region at 5.4 m/s (10.5 kts). (a) NFA predictions and (b) LIDAR measurements are respectively plotted in this figure.  The averaged wave frequency spectra for the wake centerline region within $\pm$1 m of centerline is presented at various distances aft of the transom.}
\end{center}
\end{figure}

In Figure \ref{nfa_lidar_spectra} the power spectra of the wave elevation time series for medium-resolution simulation and the LIDAR measurements are plotted for several aft transom locations (2.1, 3.1, 4.2, 5.4, 6.6, 8.0, 9.6, and 11.5 meters).  These spectra have been averaged over the centerline region within $\pm$1 m.  The frequency distribution of the predictions is of similar form to the field measurements.  However, the simulations begin at breaking onset (2.1 m aft) with more spectral energy than the field experiments (as was observed for both the predicted mean and the RMS free-surface elevations).  The predicted spectral levels then attenuate more rapidly downstream than the field measurements.  A dominant frequency at 1.3 Hz is observed for all aft transom locations for the predictions, which is above the dominant frequency of the field data (1.0 Hz), but the peak spectral amplitudes appear to be well-predicted.

%
%

\subsubsection{NFA Predictions Discussion}
\enlargethispage*{12pt}

NFA predicts the location of the onset of breaking well.  NFA predicts slightly greater mean wave heights than both the model-scale and full-scale measurements. The overprediction by NFA is significantly less at 9.3 m/s (18 kts), a dry transom with less turbulence, than for the 5.4 m/s (10.5 kt) case - suggesting that NFA does not accurately model the amount of dissipation due to turbulence caused by the breaking observed in this region. At full-scale NFA successfully captures the trend and peak locations for wave elevation, RMS, and spectral distribution,  but over-predicted the initial magnitudes, and exhibited much greater attenuation aft.  The transom was dry in the 5.4 (10.5 kt) simulation.  It may be that both the over-prediction and the rapid attenuation of the wave amplitude is partially due to the dry transom prediction.

Appendages and propellers were not included in the NFA simulations, so their absence may have contributed to the dry transom result.   The effects of the finite-difference operators that are used in NFA also requires investigation.  The predicted RMS fluctuations and the dominant frequency peak are high along the centerplane where symmetry is imposed in the NFA simulations.  Future work should investigate whether no-symmetry boundary conditions would alleviate these effects.

    %
    %

\subsection{CFDship-Iowa-V.4 Predictions} \label{subsec:CFDShip-Iowa}

The objective of this study is to evaluate the capability of CFDship-Iowa-V.4 version 4 to predict unsteady transom flows of Athena R/V in both model- and full-scale Reynolds numbers.  This objective is in contrast to the grid-density sensitivity performed in the NFA study.

The general-purpose solver CFDShip-Iowa-V.4 solves the unsteady Reynolds averaged Navier-Stokes (RANS) or detached eddy simulation (DES) equations in the liquid phase of a free-surface flow \cite{Carrica07}. The free surface is captured using a single-phase level-set method and the turbulence is modeled by isotropic or anisotropic turbulence models. Numerical methods include: advanced iterative solvers, second and higher-order finite-difference schemes with conservative formulations, parallelization based on a domain decomposition approach using the message passing interface (MPI), and dynamic overset grids for local grid refinement and large amplitude motions.

%
%
\begin{figure} [h]
\centering
\includegraphics[width=.8\columnwidth]{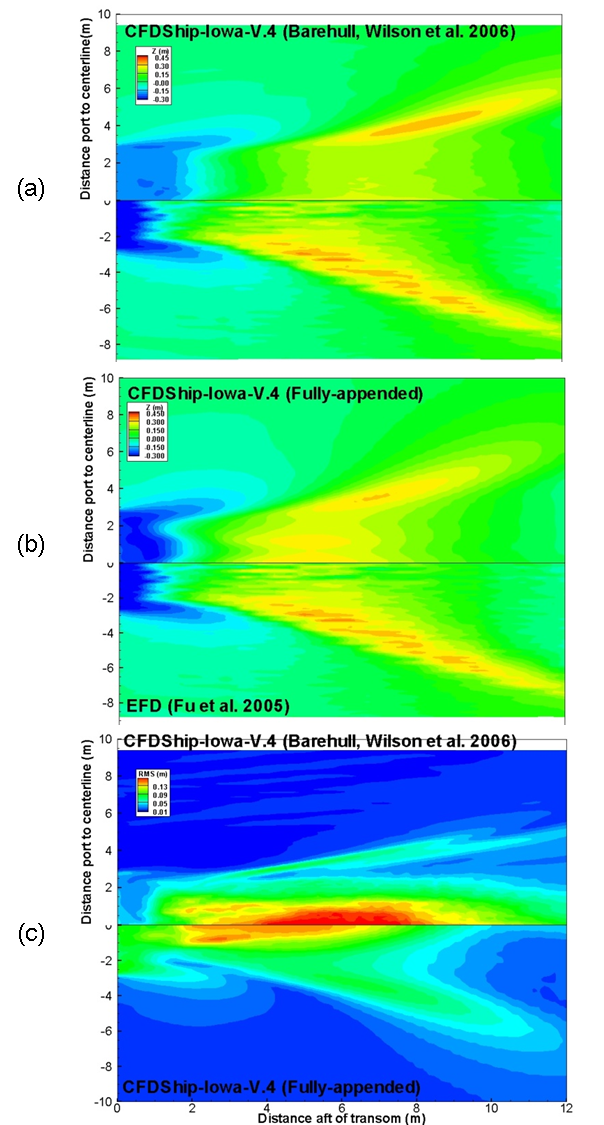}
\caption{\label{CFD_Ship_Fig_1} Model-scale results for: (a) mean free-surface elevation for bare-hull predictions and measurements, (b) fully-appended prediction of mean free surface is compared with model-scale bare-hull measurements, and (c) the RMS of the free-surface wave elevation is compared for the bare hull and fully-appended simulations.}
\end{figure}

%
%
\begin{table*} [t!]
\caption{\label{CFDShip_Table1} Simulations using CFDShip-Iowa-V.4 for fully-appended Athena at 10.5 knot.}
\begin{center}
\includegraphics[width=.9\textwidth]{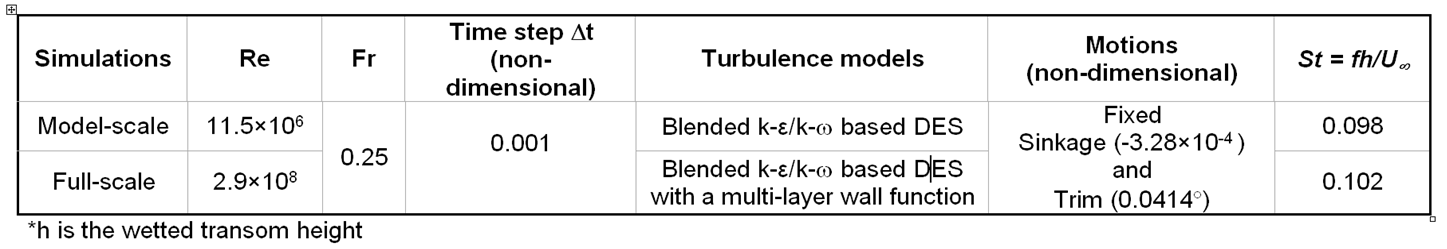}
\end{center}
\end{table*}

%
%
\begin{table*}
\caption{\label{CFDShip_Table2} Fully-Appended Athena Fine Grid Particulars.}
\begin{center}
\includegraphics[width=.9\textwidth]{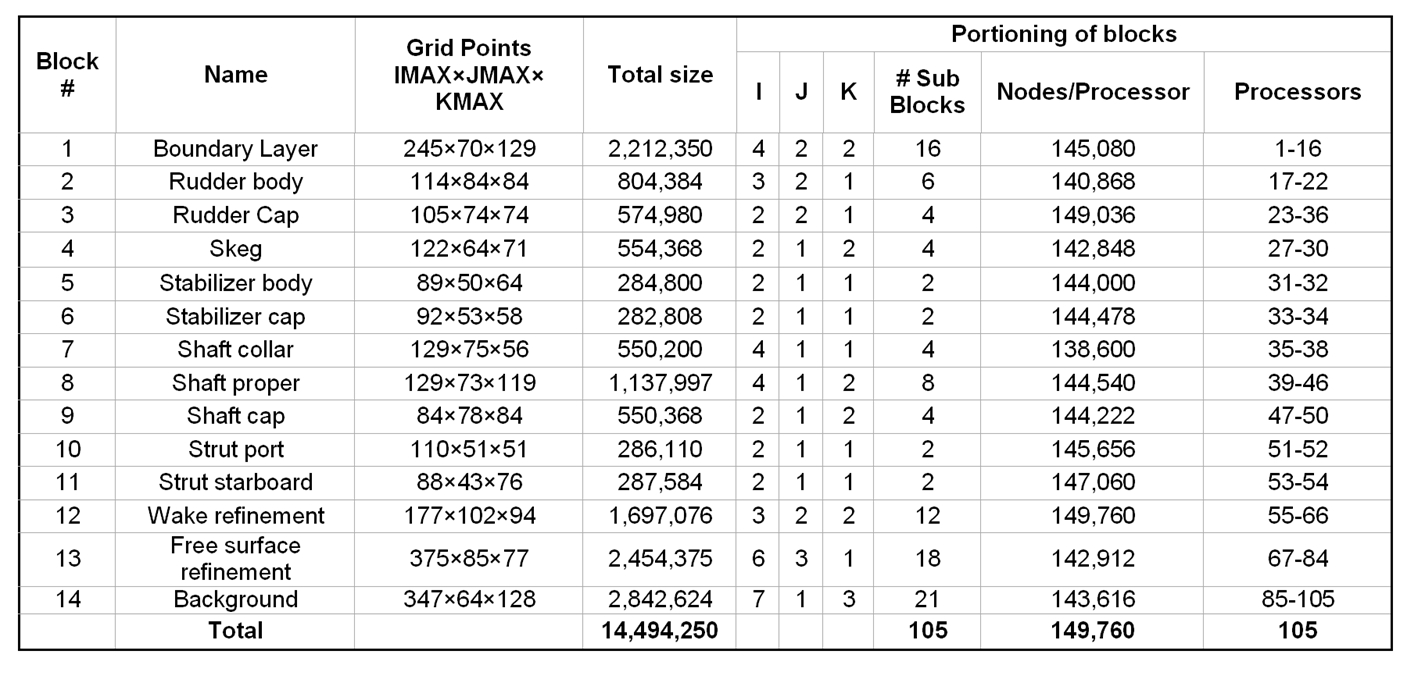}
\end{center}
\end{table*}

Simulations using CFDShip-Iowa-V.4 are summarized in Table \ref{CFDShip_Table1}. The ship simulations were performed for fixed sinkage and trim. Sinkage and trim values were established based upon previously predicted values of sinkage and trim obtained for the bare-hull geometry at the same Froude number. The grid consists of 14.5 M grid points, portioned into 105 blocks to run on parallel processors (Table \ref{CFDShip_Table2}). For convection terms in momentum, a hybrid scheme was applied, which uses third-order upwind biased for the separation region and switches to second-order very close to the wall. A second-order Euler backward difference scheme is used for the time derivative. A blended k-omega based DES model was used for turbulence modeling. Model-scale simulation was performed using a near-wall turbulence model, whereas a multi-layer wall-function (y+=100) was used for the full-scale calculation.

Simulations were performed for six ship lengths (i.e. 52 seconds). Averaging of the free-surface elevation and spectra were obtained using the last two ship lengths of time (i.e, 2000 samples), which was sufficient enough as evident from the running mean. The sampling period also corresponds to 15 turn-over times of the transom vortex shedding.

\subsubsection{Model-Scale Predictions and Comparisons}

\begin{figure}
\centering
\includegraphics[width=.9\columnwidth]{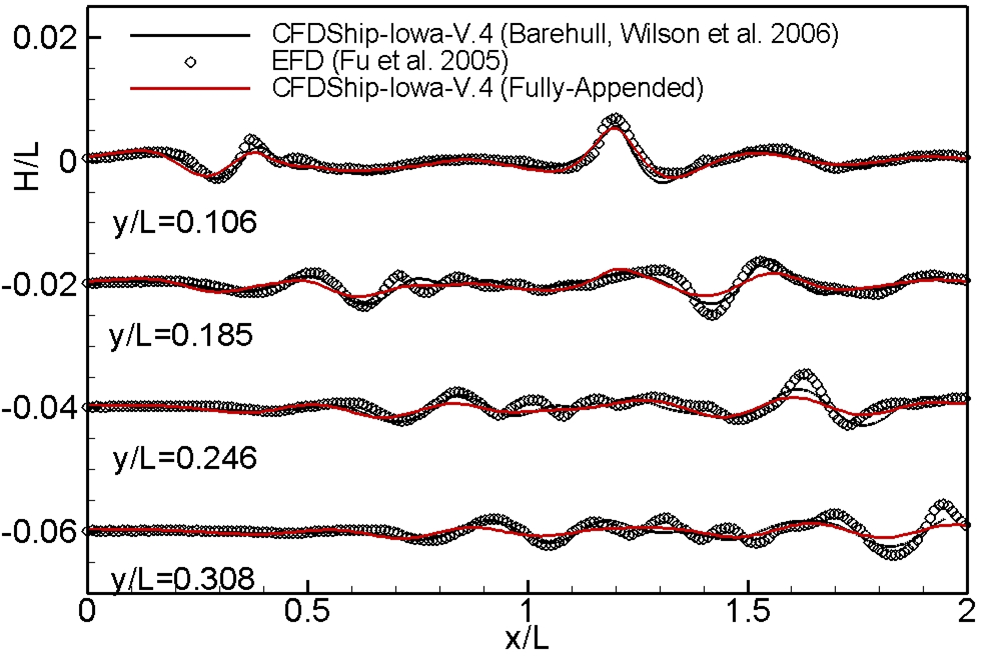}
\caption{\label{CFD_Ship_Fig_2} Wave cuts at different spanwise positions at model-scale using CFDShip-Iowa-V.4 (bare hull and fully-appended geometries) are compared with the EFD data .}
\end{figure}

In the absence of model-scale data for the fully-appended Athena, model-scale simulation results are compared with model-scale measurements and predictions for the bare-hull geometry. Figures \ref{CFD_Ship_Fig_1}a and \ref{CFD_Ship_Fig_1}b compare the mean free-surface wave elevation obtained from the fully-appended geometry with the previous bare-hull results \cite{Wilson06b} and the bare-hull data \cite{Fu05}. Figure \ref{CFD_Ship_Fig_1}c compares the predicted RMS of the free-surface wave elevation for the bare hull and fully-appended hull. In Figure \ref{CFD_Ship_Fig_2}, the predicted free-surface elevation wave-cut profiles are compared at several spanwise locations with model-scale data.

%
%

%
%
\begin{figure*}
\begin{center}
\begin{tabular}{rcrc}
(a) & & (b) & \vspace{-16pt}  \\
& \includegraphics[width=0.27\linewidth]{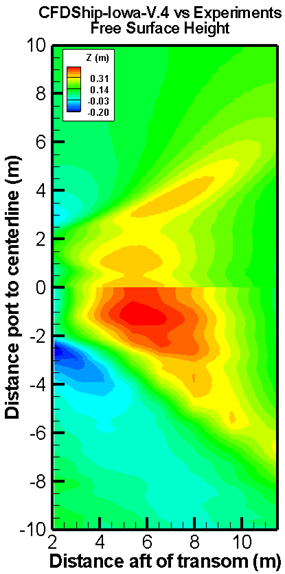}
& & \includegraphics[width=0.2889\linewidth]{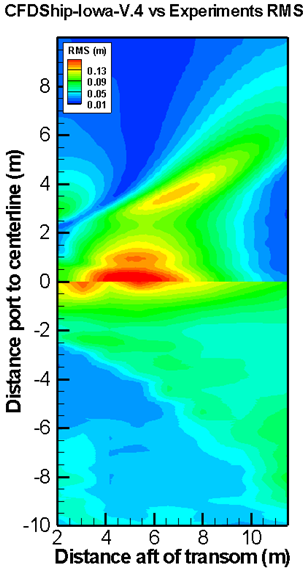} \vspace{16pt}\\
(c) & & (d) & \vspace{-16pt}  \\
& \includegraphics[width=0.27\linewidth]{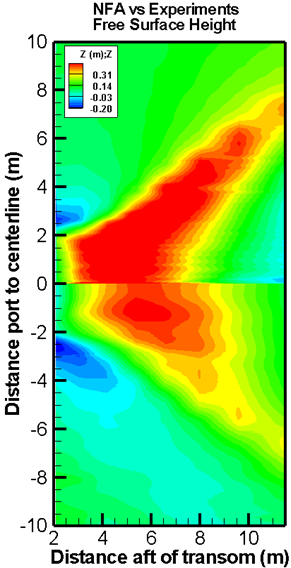}
& & \includegraphics[width=0.2673\linewidth]{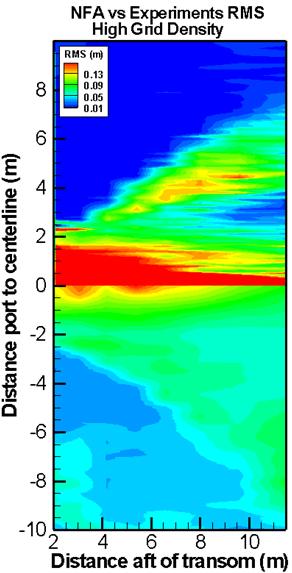}
\end{tabular}
\caption{\label{CFD_Ship_Fig_3} Simulations versus experiments in the stern region at 5.4 m/s (10.5 kts) for wave elevation mean and RMS. Lidar measurements and predictions are respectively plotted in the bottom and top halves of each plot; (a) wave elevation by CFDShip-Iowa-V.4, (b) RMS by CFDShip-Iowa-V.4, (c) wave elevation by NFA, (d) RMS by NFA.}
\end{center}
\end{figure*}

\subsubsection{Full-Scale Predictions and Comparisons}

\begin{figure}
\begin{center}
\begin{tabular}{rc}
a) & \vspace{-16pt} \\
& \includegraphics[width=0.8\columnwidth]{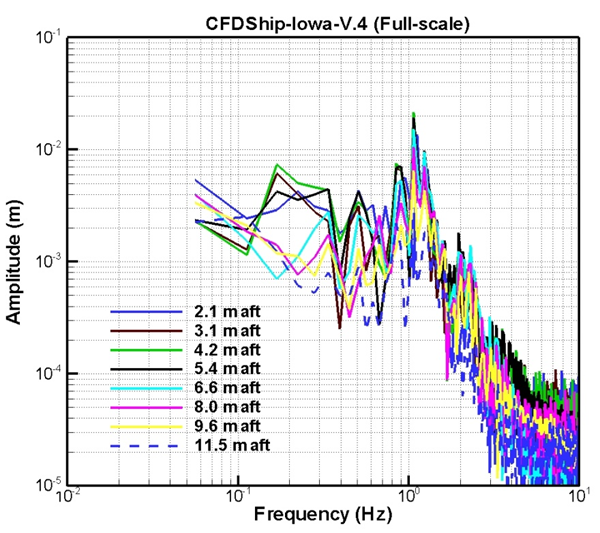} \\
(b) & \vspace{-16pt} \\
& \includegraphics[width=0.9\columnwidth]{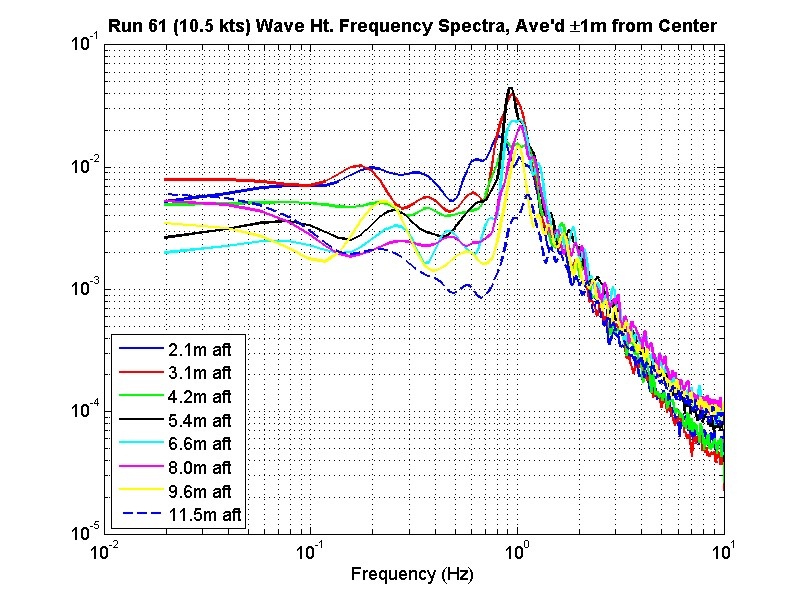}
\end{tabular}
\caption{\label{CFD_Ship_Fig_4} CFDship-Iowa-V.4 versus experiments in the stern region at 5.4 m/s (10.5 kts). (a) CFDship-Iowa-V.4 predictions and (b) LIDAR measurements are respectively plotted in this figure.  The averaged wave frequency spectra for the wake centerline region within $\pm$1 m of centerline is presented at various distances aft of the transom.}
\end{center}
\end{figure}

The CFDship-Iowa-V.4 results have been shifted vertically by 0.1 m to better match the full-scale wave elevation in the far-field.  Results are presented in dimensional units (using ship length of 46.94 m, and velocity U=5.4 m/s (10.5 kts). Figure \ref{CFD_Ship_Fig_3} shows the free-surface elevation and RMS of the free-surface wave elevation for full-scale simulations that are compared with the LIDAR measurements. NFA (unaltered) results are also presented for comparison. In Figure \ref{CFD_Ship_Fig_4} the centerline wave elevation time series FFT spectra for full-scale simulations are plotted at several aft transom locations (2.1, 3.1, 4.2, 5.4, 6.6, 8.0, 9.6, and 11.5 meters). Details of the vortical structures and associated instabilities will be presented in a separated study \cite{Bhushan08}.

%
%

\subsubsection{CFDShip-Iowa-V.4 Predictions Discussion}

The comparison between measurements, the model-scale bare-hull measurements versus the full-scale fully-appended LIDAR measurements, shows that the wave elevation peak for the appended hull is 38.5\% higher than that of bare hull.  The region of significant wave elevation for the appended hull is also much larger than that for bare hull. Unfortunately, similar observations for surface roughness can not be made, because measurements for RMS for the bare-hull configuration at model-scale are not available.

In comparing the model-scale bare-hull measurements with the bare-hull CFDShip-Iowa-V.4 simulations, the simulations predicted the mean wave crest and trough amplitudes well, but showed slightly shorter wavelength and a smaller angle for the Kelvin wave pattern.  When the model-scale bare-hull predictions of \citeasnoun{Wilson06b} were compared with the current model-scale appended-hull simulations, the current simulations predicted the same Kelvin wave angle but slightly under-predicted wave crest and trough, significantly under-predicted RMS, and exhibited a different flow pattern especially near the centerline. It is believed that the differences between the two CFD simulations is due to the effect of appendages. Another possible factor are the different grid design; \citeasnoun{Wilson06b} applied transom refinement grids that are not used in the current study.

For full-scale/appended-hull comparisons with the LIDAR measurements, CFDShip-Iowa-V.4 successfully captures the trend and peak locations for wave elevation and RMS but with under-predicted magnitudes. This deficiency of magnitudes increases downstream, and is likely due to the coarser grid resolution farther away from the ship hull.  Power spectra analysis reveals a dominant frequency at 1 Hz for all aft transom locations, which agrees well with the LIDAR spectral analysis but with amplitude under-predicted.

\begin{figure}
\centering
\includegraphics[width=.9\columnwidth]{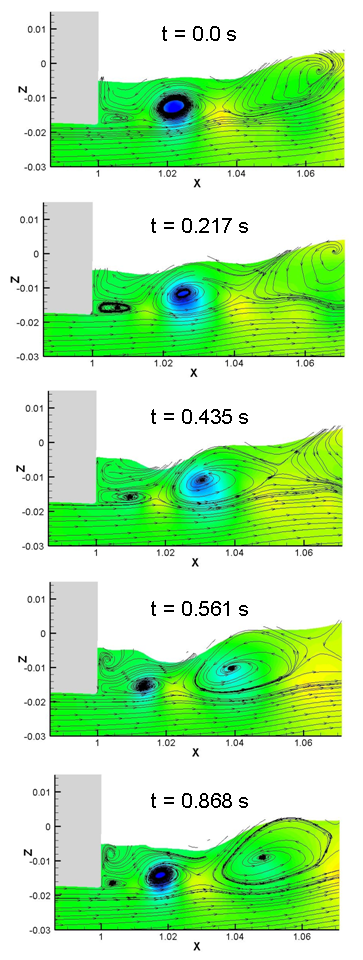}
\caption{\label{CFD_Ship_Fig_5} Phases of transom vortex shedding using CFDShip-Iowa-V.4 (full-scale simulation).}
\end{figure}

A wetted transom was observed for both model and full-scale simulations. Within the simulations, the unsteadiness of the wave elevations aft of the transom can be directly related to the Karman-like vortex shedding from the transom corner below the free surface.  Figure \ref{CFD_Ship_Fig_5} shows one complete cycle of this shedding period.  The corresponding Strouhal number (St) based on the ship velocity and wetted transom height is reported in Table \ref{CFDShip_Table1}.

Future work will be to increase the grid resolution in the transom region, add a propeller model, and evaluate the effects of fixed motions versus predicted motions.

%
%

%
%

\section{CONCLUSIONS}

An initial set of full-scale LIDAR measurements and numerical predictions of the wave-elevation topology behind a transom-sterned vessel, the R/V Athena I, have been compared and assessed in this paper. Comparisons of the mean height, surface roughness (RMS), and spectra of the breaking stern-waves at 5.4 m/s (10.5 kts), a wet-transom condition, have been made. These analyses represent (perhaps the very first) comparisons of high density in-situ data with numerical predictions of breaking stern-waves, and although preliminary, the results are reasonably encouraging.

From a naval-design perspective, where mean wave height is often the most important parameter, the numerical predictions compared reasonably well with the LIDAR data. Both codes indicated the position of the onset of breaking well, with NFA somewhat over-predicting and CDFShip-Iowa somewhat under-predicting the breaking wave heights. It is likely that, with additional effort, both codes will converge more closely to the measurements. For NFA this means the absence of the centerplane-symmetry assumption employed in this analysis, the addition of appendages, and finally the addition of propulsors. For CFDship-Iowa-V.4, this additional work includes increased grid density, and the modeling of the propulsors.

From a design-analysis perspective, where details of the flow become more significant, both numerical techniques displayed an over-suppression of the surface roughness in the late wake after the initial onset of breaking. While both codes displayed good spectral frequency distribution with respect to the measurements, it was evident that issues remain. At the onset of breaking, NFA had initially higher spectral content but attenuated too rapidly aft. CFDship-Iowa-V.4 exhibited weaker attenuation aft, but at breaking onset initially had suppressed spectral content when compared to the measurements. While in both cases the attenuation aft may be due solely to a lack of adequate grid resolution, it is also true that un-modeled turbulent processes play a very important role in this region. For example, unsteady interaction of the propulsors and appendages may set up large organized turbulent structures that affect the dissipation of surface roughness. Detailed analysis of the vortical structures and associated instabilities found in the CFD predictions needs to be performed. These vertical structure are closely connected with the time-history of transom wave elevation and resistance/pressure coefficients.  Additionally, it is likely that the significant bubble entrainment (observable at full-scale) plays an important role in the dissipation rates of turbulence in the late wake region, which neither code currently models.

The LIDAR-based wave measurement system proved itself to be a new and extremely useful tool for the understanding of wavebreaking physics in these analyses.  The LIDAR measurements have provided information and detail that have heretofore been unavailable.  However, it was necessary to de-trend the data prior to making comparisons to the predictions.  It may therefore be inappropriate to attribute all the differences between the simulations and measurements to the CFD codes.  The full-scale measurements will require additional tests for confidence before final conclusions can be properly made.

This paper represents a work in progress.  While this initial comparison of the numerical predictions and full-scale measurements has generally shown good agreement, it has also revealed areas of necessary improvement. Significantly more analysis will be required to comprehend and model the physics observed and available in this full-scale data set. Fortunately, this full-scale measurement effort represents only one small portion of the ONR Ship Wave Breaking and Bubble Wake program. It will be necessary to include data from all sources within this program to put any newly identified questions introduced by these first comparisons into context: including examination of the other sensors (QViz, capacitance probes, and Void-Fraction sensors), and the other experiments (the large Transom-Stern model test, and the Athena model tests). 
%
%

\section{ACKNOWLEDGEMENTS}

The Office of Naval Research supports this research under multiple contract vehicles. Dr. Patrick Purtell is the program manager.  The NFA work is supported in part by a grant of computer time from the DOD High Performance Computing Modernization Program (http://www.hpcmo.hpc.mil/).  The numerical simulations have been performed on the Cray XT3 at the U.S. Army Engineering Research and Development Center (ERDC).  The authors would like to thank Professor Fred Stern from IIHR-Hydroscience and Engineering, the University of Iowa, and all the experimentalists involved in the successful collection of the full-scale data.

%
%

\bibliography{27ONRwyatt}
\bibliographystyle{27onr}

%
%

\end{document}